\documentclass[a4paper, 10pt, notitlepage]{article}
\usepackage[utf8]{inputenc}
\usepackage{graphicx}
\usepackage{subcaption}
\usepackage{color}
\usepackage{url}
\usepackage{geometry}
\usepackage{caption, tabularx}
\usepackage{enumitem, ragged2e}
\usepackage[svgnames]{xcolor}

\usepackage[small]{titlesec} % Section title size, where <size> is big, medium, small, or tiny.
\usepackage{amssymb}
\usepackage{amsfonts}
\usepackage{amsthm}
\usepackage{empheq,amsmath}
\usepackage{gensymb}
\usepackage[switch]{lineno} % Line number package
\usepackage[symbol]{footmisc}

\numberwithin{equation}{section} %Numbering equations by section (to enumerate by subsection, use {subsection})

\theoremstyle{remark}

\renewenvironment{abstract}
{\begin{quote}
\noindent \rule{\linewidth}{.5pt}\par{\bfseries \abstractname.}}
{\medskip\noindent \rule{\linewidth}{.5pt}
\end{quote}
}

\begin{document}
\begin{center}
\large{\textbf{LOW-EFFOURTH: A COMPUTATIONAL FRAMEWORK FOR GENERATING AND STUDYING MULTILEVEL MODEL ENSEMBLES IN LOW-DIMENSIONAL SYSTEMS}}\\
\end{center}

\begin{center}
F. DE MELO VIR\'ISSIMO$^{1,}$\footnote{Corresponding Author. e-mail: f.de-melo-virissimo@lse.ac.uk.}\\
\end{center}

\begin{center}
{\small \it $^1$Grantham Research Institute on Climate Change and the Environment, London School of Economics and Political Science, London, WC2A 2AE, United Kingdom}\\
\end{center}

\begin{center}
     \textbf{NOTE: THIS IS A NON-PEER REVIEWED PREPRINT SUBMITTED TO ARXIV}
\end{center}

%\begin{linenumbers}
{\small
\begin{abstract}
This paper introduces Low-EFFourth (LEF4), a MATLAB-based computational framework designed for generating and studying multilevel model ensembles in continuous dynamical systems. Initially developed to address questions in climate modelling, LEF4 can also be used in other disciplines such as epidemiology, economics, and engineering, with minimal modifications to the code. The framework provides an efficient and flexible approach for investigating uncertainties arising from initial conditions, model parameters, numerical methods, and model formulation. This preprint serves as a formal reference for the LEF4 codebase and provides a concise technical and conceptual overview of its purpose, structure, applications and development pipeline.
\\
\end{abstract}

\begin{center}
\end{center}

\begin{center}
\end{center}
}

\newpage

\sloppy

%%%%%%%%%%%%%%%%%%%%%%%%%%%%%%%%%%%%%%%%%%%%%%%%%%%%%
% SECTION 1 - INTRODUCTION
%%%%%%%%%%%%%%%%%%%%%%%%%%%%%%%%%%%%%%%%%%%%%%%%%%%%%
\section{Introduction}
The study of complex systems in fields such as climate modelling, epidemiology, and economics often involves ensembles of simulations to account for uncertainties. These uncertainties may arise from a variety of sources, such as the initial state of the system and its parameters. But in many cases, running ensembles can be computationally expensive, requiring careful design and management that balances its core features such as size (number of simulations), shape (distribution of uncertainties) and length of simulation.

Low-dimensional systems offer a practical alternative for studying uncertainties and ensemble design in complex models. These systems are simplified representations that capture key dynamics of more complex models but with significantly reduced computational costs, offering a computationally efficient way to investigate ensemble dynamics, uncertainty propagation, and the effects of different types of perturbations. 

With that in mind, I developed Low-EFFourth (hereafter LEF4)~\cite{dmv2025:LEF4zenodo}. LEF4 is a MATLAB-based computational framework designed to generate and analyse large ensembles in continuous dynamical systems, being particularly efficient for low-dimensional systems. The framework is capable of running ensembles at four distinct levels: initial condition, parametric, multi-model, and multi-numerical. By providing a simple interface for specifying model parameters, numerical methods, and ensemble types, LEF4 allows users to run simulations with minimal effort and computational resources.

LEF4 was initially developed to address questions in climate modelling~\cite{dmv2023:lowdimDS,dmv2024:nonautonomous,dmv2025:bams}, where large ensembles are used to explore uncertainties in projections of future climate. However, its design makes it equally applicable (with minimal modifications) to a wide range of domains such as epidemic forecasting, economic modelling, engineering, and others where similar challenges of uncertainty quantification arise. The framework can handle both deterministic and stochastic models and offers various strategies for uncertainty quantification (UQ). Furthermore, LEF4 is can be used as an educational tool, being suitable for use in undergraduate and postgraduate classrooms to teach students about ensemble modelling in different disciplines.

%%%%%%%%%%%%%%%%%%%%%%%%%%%%%%%%%%%%%%%%%%%%%%%%%%%%%
% SECTION 2 - USING LEF4
%%%%%%%%%%%%%%%%%%%%%%%%%%%%%%%%%%%%%%%%%%%%%%%%%%%%%
\section{Using Low-EFFourth}

To get started with LEF4, users should visit the Zenodo repository~\cite{dmv2025:LEF4zenodo}, where the source file \texttt{LowEFFourth.zip} can be downloaded. An optional file, \texttt{DataAnalysisCode.zip}, containing a series of codes for postprocessing and data analysis is also available for download. Users should then refer to the \texttt{README.txt} file, which includes a detailed step-by-step guide on how to set up and use the both LEF4 and the data analysis code. The framework requires version 2023b of MATLAB and the \texttt{cmocean}\footnote{Available for download here: \url{https://www.mathworks.com/matlabcentral/fileexchange/57773-cmocean-perceptually-uniform-colormaps}} colormap~\cite{thyng2016:colourmap} for data visualisation. Once LEF4 is installed, users can run ensemble simulations, analyse the results, and visualise the data using the provided post-processing tools.

The core functionality of LEF4 is controlled through the MATLAB function \texttt{EnsembleGenFunc()}, which generates ensembles based on the specified model, numerical method, ensemble type, and other parameters. A simple user interface is provided through the \texttt{MainEnsembleProgram.m} file, which allows users to specify key parameters and run simulations with ease. A short overview of LEF4's structure is provided in the next section.

%%%%%%%%%%%%%%%%%%%%%%%%%%%%%%%%%%%%%%%%%%%%%%%%%%%%%
% SECTION 3 - LEF4 STRUCTURE
%%%%%%%%%%%%%%%%%%%%%%%%%%%%%%%%%%%%%%%%%%%%%%%%%%%%%
\section{Framework Structure}

The LEF4 framework is implemented entirely in MATLAB and organised into a modular directory structure to facilitate ease of use, modification, and expansion. This section provides a technical overview of the main components included in version v0.0 of the framework.

\subsection*{Root Directory: \texttt{/LowEFFourth}}

\begin{itemize}
  \item \texttt{MainEnsembleProgram.m}: A user-friendly driver script containing all parameter options for an ensemble simulation. This file is used to initialise and launch runs, specifying model settings, ensemble type, solver choice, and output preferences.
\end{itemize}

\subsection*{Subdirectories and Key Contents}

\begin{description}
  \item[\texttt{/AuxiliaryFunctions}.] Contains the core functional scripts:
  \begin{itemize}
    \item \texttt{EnsembleGenFunc.m}: Primary function executing the ensemble run.
    \item \texttt{EnsembleProgRunLOGFILE.m}: Creates a logfile documenting each simulation.
    \item \texttt{AddToLOGFILE.m}: Appends run-specific metadata to the logfile.
  \end{itemize}

  \item[\texttt{/Models}.] Includes nine predefined models:
  \begin{itemize}
    \item Five test models (e.g. linear growth, harmonic oscillator).
    \item Four physical models, including coupled systems such as Stommel 61 and Lorenz 84.
  \end{itemize}

  \item[\texttt{/NumericalMethods}.] Hosts 11 numerical schemes, including:
  \begin{itemize}
    \item Classical 4th-order Runge-Kutta (RK4).
    \item Adams-Bashforth methods of various orders.
    \item Note: Some solvers are experimental or partially implemented.
  \end{itemize}

  \item[\texttt{/OutputFiles}.] Target directory for simulation outputs.
  \begin{itemize}
    \item Supports both NetCDF (\texttt{.nc}) and MATLAB's native format (\texttt{.mat}).
  \end{itemize}
\end{description}

\subsection*{Usage Pipeline}

\begin{enumerate}
  \item Set ensemble parameters in \texttt{MainEnsembleProgram.m}, including:
  \begin{itemize}
    \item Model and solver selection
    \item Ensemble type and size
    \item Time step, duration, and output frequency
    \item Output format and file name
  \end{itemize}
  
  \item Run the script, which internally calls \texttt{EnsembleGenFunc.m}:
  \begin{itemize}
    \item Perturbs inputs (e.g. initial conditions or parameters)
    \item Executes the ensemble simulation
    \item Writes output data and metadata
  \end{itemize}
\end{enumerate}

\subsection*{Post-Processing and Analysis (Optional)}

As previously menstioned, users can additionally install the optional \texttt{DataAnalysisCode.zip}, which provides post-processing and analysis tools:

\begin{itemize}
  \item \textbf{Postprocessing:} e.g. generating 1-year time-averaged output files.
  \item \textbf{Visualisation:} e.g. heatmaps, distributions, time series, phase spaces.
  \item \textbf{Statistics:} e.g. Kolmogorov-Smirnov, Jensen-Shannon divergence.
\end{itemize}

These tools require the \texttt{cmocean} colormap package~\cite{thyng2016:colourmap}.

\subsection*{Platform Compatibility}

\begin{itemize}
  \item Developed on macOS with MATLAB 2023b.
  \item Compatible with MATLAB 2019b and later (minor tweaks may be required).
  \item Windows users should replace forward slashes (\texttt{/}) with backslashes (\texttt{\textbackslash}) in path definitions.
\end{itemize}

\subsection*{Extensibility}

Users may integrate new models or solvers by modifying:
\begin{itemize}
  \item \texttt{EnsembleGenFunc.m}: Add new \texttt{case} blocks to include additional dynamics or solvers.
  \item \texttt{EnsembleProgRunLOGFILE.m}: Extend logging capabilities to reflect new configurations.
\end{itemize}

These extensions will be automated in future releases to facilitate plug-and-play extensibility.

%%%%%%%%%%%%%%%%%%%%%%%%%%%%%%%%%%%%%%%%%%%%%%%%%%%%%
% SECTION 4 - EXAMPLES
%%%%%%%%%%%%%%%%%%%%%%%%%%%%%%%%%%%%%%%%%%%%%%%%%%%%%
%\section{Examples}

%%%%%%%%%%%%%%%%%%%%%%%%%%%%%%%%%%%%%%%%%%%%%%%%%%%%%
% SECTION 5 - FUTURE DEVELOPMENTS
%%%%%%%%%%%%%%%%%%%%%%%%%%%%%%%%%%%%%%%%%%%%%%%%%%%%%
\section{Future Developments}
While LEF4 is already operational and ready for use, an upcoming version (v1.0) is underway, which will bring several improvements:

\begin{itemize}
  \item \textbf{Ease of Extensibility:} It will allow users to add new models or solvers without modifying the core code. This will make LEF4 even more flexible and user-friendly.
  \item \textbf{Multi-Model and Multi-Numerical Ensembles:} It will support the automatic generation of multi-model and multi-numerical ensembles, allowing for more comprehensive uncertainty quantification across models and numerical schemes.
  \item \textbf{Performance Optimisation:} It will feature an optimised code, improving performance of LEF4, reducing computation time and improving scalability for large ensembles.
  \item \textbf{Integration with Version Control:} Version v1.0 will be shared via GitHub, enabling version control and fostering collaboration within the scientific community.
%  \item \textbf{Educational Features:} Beyond the upgrades above, LEF4 will continue to evolve as an educational tool, with additional documentation, tutorials, and visualisation tools aimed at supporting students and researchers.
\end{itemize}

%%%%%%%%%%%%%%%%%%%%%%%%%%%%%%%%%%%%%%%%%%%%%%%%%%%%%
% ACKNOWLEDGEMENTS
%%%%%%%%%%%%%%%%%%%%%%%%%%%%%%%%%%%%%%%%%%%%%%%%%%%%%
\section*{Acknowledgements}
F.d.M.V. acknowledge partial support received from a UK Natural Environment Research Council grant (ODESSS, agreement number NE/V011790/1) for an early development of this work.

%%%%%%%%%%%%%%%%%%%%%%%%%%%%%%%%%%%%%%%%%%%%%%%%%%%%%
% DATA AVAILABILITY
%%%%%%%%%%%%%%%%%%%%%%%%%%%%%%%%%%%%%%%%%%%%%%%%%%%%%
\section*{Data availability}
LEF4 and its documentation are freely available for download on Zenodo~\cite{dmv2025:LEF4zenodo}.

%%%%%%% BIBLIOGRAPHY
\bibliographystyle{IEEEtran} % Bibliography style alpha
\bibliography{2025deMeloVirissimo_Low-EFFourth}  % Bibliography file 'biblio.bib'

\end{document}